\documentclass[twocolumn,showpacs,showkeys,amsmath,amssymb]{revtex4}

\usepackage{graphicx}% Include figure files
\usepackage{bm}% bold math

\def\slashchar#1{\setbox0=\hbox{$#1$}
   \dimen0=\wd0 \setbox1=\hbox{/} \dimen1=\wd1
   \ifdim\dimen0>\dimen1 \rlap{\hbox to \dimen0{\hfil/\hfil}} #1
   \else  \rlap{\hbox to \dimen1{\hfil$#1$\hfil}} / \fi}

\begin{document} 
\title{Pion light-cone wave function and pion distribution amplitude in the 
Nambu--Jona-Lasinio model} 

\author{Enrique Ruiz Arriola} \email{earriola@ugr.es}
\affiliation{Departamento de F\'{\i}sica Moderna, Universidad de
Granada,E-18071 Granada, Spain} \author{Wojciech Broniowski} 
\email{Wojciech.Broniowski@ifj.edu.pl} \affiliation{The
H. Niewodnicza\'nski Institute of Nuclear Physics, PL-31342 Krak\'ow,
Poland}

\date{12 August 2002, ver. 2} 

\begin{abstract}
We compute the pion light-cone wave function and the pion quark
distribution amplitude in the Nambu--Jona-Lasinio model. We use the
Pauli-Villars regularization method and as a result the distribution
amplitude satisfies proper normalization and crossing properties. In
the chiral limit we obtain the simple results, namely $\varphi_\pi
(x)=1 $ for the pion distribution amplitude, and $ \int d^2 k_\perp
\Psi_\pi(x,\vec k_\perp ) k_\perp^2 = \langle \vec k_\perp^2
\rangle = -M \langle \bar u u \rangle / f_\pi^2$ for the second 
moment of the pion light-cone wave function, where $M$ is the
constituent quark mass and $f_\pi$ is the pion decay constant. After
the QCD Gegenbauer evolution of the pion distribution amplitude 
good end-point behavior is recovered, and
a satisfactory agreement with the analysis of the experimental data
from CLEO is achieved. This allows us to determine the momentum scale
corresponding to our model calculation, which is close to the value $Q_0=313$~MeV
obtained earlier from the analogous analysis of the pion parton distribution function.
The value of  $\langle \vec k_\perp^2
\rangle$ is, after the QCD evolution, around $(400~{\rm MeV})^2$. In
addition, the model predicts a linear integral relation between the
pion distribution amplitude and the parton distribution function of
the pion, which holds at the leading-order QCD evolution.
\end{abstract} 



\maketitle

\section{Introduction} 
The study of high-energy exclusive processes~\cite{BL80} provides a
convenient tool of learning about the quark substructure of hadrons.  In
this limit the total amplitude factorizes into a hard contribution,
computable from perturbative QCD, and a soft matrix element which
requires a non-perturbative treatment. From the point of view of
chiral symmetry breaking a particularly interesting process is
provided by the $\gamma^* \to \gamma^* \pi^0 $ transition form
factor. For real photons its normalization is fixed by the anomalous
breaking of chiral symmetry by the $\pi^0 \to \gamma \gamma $
decay. In addition, in the limit of large photon virtualities,
factorization allows us to define the leading-twist pion distribution
amplitude as a low energy matrix element whose normalization is fixed
by the pion weak-decay constant, a spontaneous chiral symmetry
breaking feature of the QCD vacuum. It seems obvious that such a
process offers a unique opportunity not only to learn about the
interplay between high and low energies, but also to understand the
relation between the spontaneous and the anomalous chiral symmetry
breaking. Radiative logarithmic corrections to the pion distribution
amplitude (PDA) can be easily implemented through the QCD evolution
equations~\cite{BL79,Mu95}, which yield for $Q^2 \to \infty$ the
asymptotic wave function of the form $\varphi_\pi(x,\infty) = 6 x(1-x)$. 
Moreover, the pion transition form factor has been
measured by the CELLO~\cite{cello} and, recently, the 
CLEO collaborations~\cite{CLEO98}. A theoretical
analysis of PDA based on these data and light-cone sum rules has been
undertaken~\cite{SY00}, showing that at $Q=2.4$~GeV PDA is neither
asymptotic, nor possesses the double-hump structure \cite{CZ84} proposed in early
works \cite{E791} \footnote{A recent direct measurement of PDA via the di-jet production 
by the E791 collaboration \cite{E791} shows that at scales $Q \simeq 5-6$ GeV, the 
possible admixture of the Chernyak-Zhitnitsky wave function is rather small}.

The pion distribution amplitude has been evaluated with QCD sum
rules~\cite{MR,RR,BJ97,BM,BMS1,BMS2}, 
in standard~\cite{DP00} (only the second $\xi$-moment) and
transverse lattice approaches~\cite{Da01,BS01,BD02}, and in chiral quark
models~\cite{ET,PP97,PP99,ADT00,He00,He01,PR01,ADT01,ADT01a,DVY02,Do02}. In
chiral quark models the results are not always compatible to each
other, and even their interpretation has not always been the
same. While in same cases there are problems with chiral symmetry and
proper normalization~\cite{PP97,PP99,PR01}, in other cases
\cite{ADT00,He00,He01,PR01,ADT01,DVY02,Do02} it is not clear how to
associate the scale at which the model is defined, necessary to define
the starting point for the QCD evolution. Nevertheless, there is a
precise way to identify the low energy scale, $Q_0$, at which the
model is defined, namely the one at which the quarks carry $100\%$ of
the total momentum~\cite{JG80,Ja85}. The fact that several
calculations~\cite{PP97,PP99,He00,He01,PR01,ADT01,DVY02} produce a PDA
strongly resembling the asymptotic form suggests that their working
scale is already large, and the subsequent QCD evolution becomes
unnecessary, or numerically insignificant.  This also tacitly assumes
that these models already incorporate the QCD radiative corrections.
  
In the present paper we compute the pion distribution amplitude and
the pion light-cone wave function within the Nambu--Jona-Lasinio (NJL)
model~\cite{NJL61,NJL} in a semibosonized form using the Pauli-Villars
(PV) regularization method~\cite{PV49}. This method has been
introduced in Refs.~\cite{Ru91,SR92} in the context of chiral
perturbation theory, as well as for chiral solitons. From the point of
view of the NJL model the study of exclusive processes becomes
interesting in its own right. Although factorization holds beyond
doubt in QCD, it is far from obvious that any of the regularization
schemes used to make a low-energy model well defined is compatible
with factorization. In addition, we want to determine what is the
low-energy scale, $Q_0$, the model corresponds to.  Here we obtain it with
help of the analysis of PDA and compare it to the $Q_0$ obtained in deep
inelastic scattering (DIS) from the corresponding parton distribution
function of the pion (PDF).

To a large extent our treatment of PDA parallels the calculation of PDF 
carried out in previous works~\cite{DR95,WRG99,DR02}. There,
it has been argued that for inclusive processes, such as in deep inelastic
scattering, by far the most convenient regularization scheme is the
Pauli-Villars (PV) method. Such a regularization allows
the extraction of the leading-twist contribution to the forward virtual
Compton amplitude which possesses proper support and normalization. The
relevance of regularization in chiral quark models should not be
underestimated; it is not evident what is the most convenient way to
cut-off high energies in such a way that most features of QCD are
retained. Those include chiral symmetry, gauge invariance, and scaling
properties. The main outcome of the calculation presented in
Ref.~\cite{DR95} was that, at the scale $Q_0$ at which the model is
defined, the valence PDF is a constant equal to one,
\begin{equation}
q(x,Q_0) = \bar q(1-x, Q_0) \equiv V_\pi(x,Q_0)/2=1.
\label{PDF1}
\end{equation} 
After QCD evolution at leading order (LO), impressive agreement with
the analysis of Ref.~\cite{SM92} at the reference scale $Q=2 {\rm
GeV}$ has been achieved.  At this scale the valence quarks carry $47
\%$ of the total momentum. This implies a rather low scale $Q_0$, as
suggested by the evolution ratio $\alpha(2 {\rm GeV})/\alpha(Q_0)=0.15
$ relevant at leading order.  For $\alpha (2 {\rm GeV}) = 0.32 $
listed in the PDG~\cite{PDG}, and for the evolution with three
flavors, this corresponds to~\footnote{In this paper we use the LO QCD evolution, where
$ \alpha(Q)= ( 4 \pi / \beta_0 ) / \log (Q^2 / \Lambda_{\rm QCD}^2
) $ with $\Lambda_{\rm QCD}=226~{\rm MeV} $ for $N_F=3$.}
\begin{equation}
Q_0=313 \, {\rm MeV}, \,\,\, \alpha(Q_0)= 2.14 
\label{Q0}
\end{equation} (see Ref.~\cite{DR95} for details). 
The low scales are confirmed by the next-to-leading (NLO) analysis of
Ref.~\cite{DR02}, with the NLO effects small compared to the LO ones
\footnote{Admittedly, one may worry that such a low scale as in
Eq. (\ref{Q0}) may invalidate the use of perturbation theory. We
hope that the correspoding value of $\alpha /(2 \pi )\sim 0.3$, which
typically is the expansion parameter, is low enough for the approach
to make sense.}. Motivated by this success, in the present paper we
investigate whether the evolution ratio and the values (\ref{Q0})
found in deep inelastic scattering are compatible with the values
extracted from a similar analysis of PDA at LO in the same model (NJL)
with the same (PV) regularization. This is the main objective of this
work.

In the NJL model PDA has already been estimated by several
authors~\cite{He00,He01,DVY02}. The work of Refs.~\cite{He00,He01}
uses the Brodsky-Lepage cut-off regularization as suggested by the
light-front quantization formalism. As a consequence, the asymptotic
form $\varphi(x,Q_0) = 6 x(1-x)$ is obtained without any additional
evolution. On the other hand, the same regularization yields PDF of
the form $x V_\pi (x,Q_0) \sim 6 x^2 (1-x)$~\cite{He01,BH99} which is
far from the asymptotic value $ x V_\pi(x,\infty)= x \delta(x) =0$. 
This is a rather puzzling result, which may have to do with
subtleties of introducing a regularization in the light-cone
quantization method (see also Ref.~\cite{BH99}). For that reason we
prefer to use a manifestly covariant formalism, where chiral symmetry
can be easily implemented in presence of the regularization.  In
Ref.~\cite{DVY02} PDA has been extracted from the transition form
factor by examining the asymptotic behavior for large photon
virtualities.  This requires introducing a regularization for an
abnormal parity process which also modifies the chiral anomaly, and
hence, for typical parameter values~\cite{BH88}, the $\pi^0 \to
\gamma \gamma $ decay rate is reduced by $40\%$ of the current algebra
value. Our approach is free of such problems.

\section{The Nambu--Jona-Lasinio model} 

For the reader's convenience we briefly review the NJL model 
in such a way that our results can be easily
stated. The SU(2) NJL Lagrangian in the Minkowski space is given by
\cite{NJL61,NJL}
\begin{eqnarray}
{\cal L}_{\rm NJL} &=&
\bar{q} (i\slashchar\partial - M_0 )q  +
{G \over 2} \left( (\bar{q}  q)^2
                        +(\bar{q}\vec \tau i \gamma_5 q)^2 \right)  
\end{eqnarray}
where $q=(u,d )$ represents a quark spinor with $N_c $ colors,
$\vec \tau $ are the Pauli isospin matrices, $ M_0 $
stands for the current quark mass, and $G $ is the coupling constant.
In the limiting case of the vanishing $M_0$ the action is invariant
under the global $SU(2)_R \otimes SU(2)_L $ transformations. With help
of bosonization, the vacuum-to-vacuum transition amplitude
in presence of external vector and axial-vector currents,$(v,a)$, can be written as
the path integral
\begin{eqnarray*} 
&& \langle 0| {\rm T} {\rm exp} \Bigl\{ i \int d^4 x
\left [ \bar q \left ({\slashchar v}+{\slashchar a} \gamma_5 \right ) q 
\right ] \Bigr\} |0 \rangle \\
&& = \int D \Sigma D \vec \Pi \, {\rm exp}\{{\rm i} S \}.
\end{eqnarray*} 
The following Dirac operators
\begin{eqnarray*} 
 i {\rm D} \, & = & i\slashchar{ \partial } - M_0 - ( \Sigma + i\gamma_5
\vec \tau \cdot \vec \Pi ) + {\slashchar v}+{\slashchar a} \gamma_5, \\ 
i {\rm D}_5 & = & -i\slashchar{ \partial } - M_0 - (
\Sigma - i\gamma_5 \vec \tau \cdot \vec \Pi ) + {\slashchar
v}-{\slashchar a} \gamma_5 ,
\end{eqnarray*} 
are introduced. The fields $(\Sigma,\vec \Pi )$ are dynamical,
internal bosonic scalar-isoscalar and pseudoscalar-isovector fields,
which after suitable renormalization can be interpreted as the
physical $\sigma$ and pion fields.  The PV-regularized normal parity 
($\gamma_5$-even) contribution to the effective action is~\cite{Ru91,SR92}
\begin{eqnarray} 
S_{\rm even} = - {i N_c \over 2} \sum_i c_i {\rm tr} \log ( {\rm D} {\rm D}_5 +
\Lambda_i^2 + i\epsilon) \nonumber \\
-{1\over 2G} \int d^4 x
( \Sigma^2 + \vec \Pi^2 ),
\end{eqnarray} 
with ${\rm tr}$ denoting the trace in the Dirac and isospin space. 
In general, we assume $n$  PV subtractions, with the conditions
$\sum_{i=0}^n c_i \Lambda_i^{2k}=0$ for $k=0, ... , n$ , and with $c_0=1$, 
$\Lambda_0=0 $. At least two subtractions ($n=2$), which is the case
used throughout this paper, are needed to regularize the quadratic
divergence. The abnormal parity ($\gamma_5$-odd) contribution to the
effective action is
\begin{eqnarray} 
S_{\rm odd} = - {iN_c \over 2} \left\{  {\rm tr} \log ( {\rm D}^2  ) - {\rm tr} \log
({\rm D}_5^2 ) \right\} 
\label{eq:abnor} 
\end{eqnarray} 
Notice that no explicit finite cut-off regularization is introduced in
the abnormal parity contribution, as demanded by a proper reproduction
of the QCD chiral anomaly. This subtle and important point has been
discussed in detail in Ref.~\cite{SR95}.

Any mesonic correlation function can be obtained from this
gauge-invariantly regularized effective action by a suitable functional
differentiation with respect to the relevant external fields. In
practice, one usually works in the formal limit large $N_c$, in other
words, at the one-quark-loop level. To fix the parameters in the
PV-regularized NJL model we proceed as usual (see, {\em e.g.},
Ref.~\cite{SR92}). The effective potential leads to dynamical chiral
symmetry breaking, thereby yielding a dynamical quark mass, $M$, and
condensates given by
\begin{eqnarray} 
\langle \bar u u \rangle = \langle \bar d d \rangle = -{M - M_0 \over
2 G } = 4 N_c M   I_2,
\end{eqnarray} 
where the quadratically-divergent integral, $I_2$, is defined as
\begin{eqnarray} 
I_2 &=& {i} \int {d^4 k \over (2\pi)^4 } \sum_i c_i {1\over (-k^2
+ M^2 + \Lambda_i^2 -i \epsilon )} \nonumber \\ &=&\frac1{(4\pi)^2}
\sum_i c_i (\Lambda_i^2+M^2) \log ( \Lambda_i^2+ M^2). \label{I2}
\end{eqnarray} 
The calculation of the relevant correlation function yields for the pion mass
\begin{eqnarray} 
m_\pi^2 ={ 2 I_2 \over F ( m_\pi^2 ) }{M_0 \over M -M_0 } .
\end{eqnarray} 
The pion weak-decay constant, $f_\pi$, and the pion-quark
coupling constant, $g_{\pi qq}$, are  given by
\begin{eqnarray} 
f_\pi &=& 4 N_c M F ( m_\pi^2 ) g_{\pi qq}, \label{fpi} \\ 
 {1\over g_{\pi q q}^2} &=& 4N_c {d\over d p^2 }
 \Bigl\{ p^2 F ( p^2 ) \Bigr\} \Big|_{p^2 = m_\pi^2},
\label{eq:deffpi} 
\end{eqnarray} 
respectively.  We have introduced the following short-hand
notation:
\begin{eqnarray} 
F (p^2) = \int_0^1 dx F (p^2, x),
\end{eqnarray} 
where, in terms of the PV-regularized one-loop integrals,
\begin{eqnarray} 
&& F (p^2, x) = -i \int {d^4 k \over (2\pi)^4 } \sum_i  \times \\
&& c_i {1\over [-k^2 -x(1-x)p^2 + M^2 + \Lambda_i^2 - i\epsilon]^2} \nonumber \\ 
&& = -\frac1{(4\pi)^2}\sum_i c_i \log\left[ M^2 + \Lambda_i^2 -x(1-x)p^2 \right]. 
\nonumber
\label{F}
\end{eqnarray}
The function $F$ in an obvious manner 
satisfies the symmetry relation $F (p^2 , x) = F (p^2, 1-x)$.  In
the case of two subtractions, and in the limit $\Lambda_1 \to
\Lambda_2 \equiv \Lambda $ used in this paper, we have $\sum_i c_i f(\Lambda_i^2) = f(0)
- f(\Lambda^2 ) + \Lambda^2 f' (\Lambda^2 ) $. 
In the numerical analysis of this paper we work in the strict chiral limit, with $M_0=0$. 
The parameters are fixed as usual; we adjust the cut-off, $\Lambda$, in order to
reproduce the physical pion weak-decay constant, $f_\pi = 93.3 $ MeV. The coupling constant, $G$,
is traded for the constituent quark mass, $M$, which remains the only free parameter of the model. 
In our study of the pion light-cone wave function we use two sets, which cover the 
range used in other phenomenological applications of the model: $M = 280$ MeV,
$\Lambda=871$ MeV (case of Ref. \cite{DR95}), and $M = 350$ MeV,
$\Lambda=770$ MeV. These give
the quark condensate equal to $\langle \bar u u + \bar d d \rangle = - ( 290 {\rm MeV} )^3$ and 
$- ( 271 {\rm MeV} )^3$, respectively. 
As we shall see, the results are insensitive to the choice of parameters.

\section{Pion light-cone wave function and pion distribution amplitude} 

The pion light-cone wave function (the axial-vector component) is defined
as the low-energy matrix element~\footnote{The light-cone coordinates
are defined as $ k^\pm = k^0 \pm k^3 $ and $d^4 k = \frac12 dk^+ dk^-
d^2 \vec k_\perp $}
\begin{eqnarray}
&&\Psi_{\pi } (x, \vec k_\perp ) = -\frac{{i} \sqrt{2} }{4\pi
f_\pi} \int d \xi^- d^2 \xi_\perp e^{{i} (2x-1) \xi^- p^+ -
\xi_\perp \cdot k_\perp } \nonumber 
\times \\ && \langle \pi^+ (p) | \bar u (\xi^- ,
\xi_\perp) \gamma^+ \gamma_5 d(0) | 0 \rangle .
\label{eq:pda_def}
\end{eqnarray} 
where $p^\pm = m_\pi $ and $\vec p_\perp= 0$.  The pion
distribution amplitude is defined as
\begin{eqnarray}
\varphi_{\pi } (x) = \int d^2 k_\perp \Psi_{\pi } (x, \vec
k_\perp) 
\end{eqnarray} 
Formally, in the momentum space, 
Eq.~(\ref{eq:pda_def}) corresponds to integration over
the quark momenta in the loop integral used in
the evaluation of $f_\pi$, but with $k^+ = p^+ x = m_\pi x $ 
and $k_\perp$ fixed. Thus, with the PV method
and after working out the Dirac traces, we have to compute
\begin{eqnarray}
&& \Psi_{\pi } (x, \vec k_\perp ) = -\frac{ 2 i N_c M  g_{\pi qq} }{
f_\pi} \int \frac{dk^+ dk^- }{(2\pi)^4 } \times \nonumber 
\\ && \frac{\delta \left(
k^+ - x p^+ \right)}{m_\pi x(1-x) }
   \sum_j  c_j \times  \\&& \frac1{k^- - m_\pi + \frac{ \vec k_\perp^2 + M^2 + \Lambda_j^2 +
i 0^+ }{m_\pi (1-x) } } \, \frac1{k^- - \frac{ \vec k_\perp^2 + M^2 +
\Lambda_j^2 + i 0^+}{m_\pi x } }\nonumber,
\end{eqnarray}
where the location of the poles in the $k^-$ variable has been explicitly
displayed. Evaluating the $k^-$ integral gives the pion LC wave
function in the NJL model with the PV regularization:
\begin{eqnarray} 
\Psi_{\pi } (x, k_\perp)&=& \frac{4 N_c M g_{\pi qq} }{16\pi^3 f_\pi}
\sum_j  c_j \times \nonumber \\
&&\frac1{k_\perp^2 + \Lambda_j^2 + M^2-x(1-x)m_\pi^2 } .
\end{eqnarray} 
The function is properly normalized, 
\begin{equation} \int d^2 k_\perp dx \Psi_{\pi }
(x, k_\perp)=1, \end{equation} 
and satisfies the crossing relation 
\begin{equation}
\Psi_{\pi}
(x , \vec k_\perp ) = \Psi_{\pi} (1-x , \vec k_\perp ). 
\end{equation} 
For $m_\pi \neq 0 $ it is non-factorizable in the $k_\perp$ and $x$ variables.
Integrating with respect to $k_\perp$ yields the pion distribution
amplitude,
\begin{eqnarray} 
\varphi_{\pi} (x)&=& 4 N_c M F ( m_\pi^2 , x ) \frac{g_{\pi
qq}}{f_\pi}.
\end{eqnarray} 
The crossing property, $\varphi_\pi(x)=\varphi_\pi(1-x)$ follows trivially, 
and Eq. (\ref{fpi}) gives the correct normalization, namely
$\int dx \, \varphi_{\pi} (x)=1$.  

As a consequence of the PV condition with two subtractions one has,
for large $k_\perp$,
\begin{eqnarray} 
\Psi_{\pi} (x, k_\perp)& \to & \frac{4 N_c M^2}{16\pi^3
f_\pi^2} \frac{ \sum_i c_i \Lambda_i^4 }{k_\perp^6},
\end{eqnarray} 
which gives a finite normalization and a finite second transverse
moment, 
\begin{eqnarray} 
\langle k_\perp^2 \rangle &=& \int d^2 k_\perp \int_0^1 dx \, \Psi_{\pi}
(x, k_\perp) k_\perp^2
\end{eqnarray} 
In the chiral limit, $m_\pi = 0$, one can use the Goldberger-Treiman
relation for the constituent quarks, $ g_{\pi qq} f_\pi = M$. Then 
$f_\pi^2 = 4N_c M^2 F(0)$, which gives the very simple formulas
\begin{eqnarray} 
\Psi_{\pi} (x, k_\perp)&=& \frac{4 N_c M^2}{16\pi^3 f_\pi^2} \sum_i
c_i \frac1{k_\perp^2 + \Lambda_i^2 + M^2 } ,
\label{eq:lc0} 
\\ \varphi_\pi (x) &=& 1 ,\label{eq:pd0} \\ \langle \vec k_\perp^2
\rangle &=& -\frac{M \langle \bar u u \rangle }{f_\pi^2}.
\label{eq:kp0} 
\end{eqnarray} 
In the chiral limit $\Psi_\pi ( x, \vec k_\perp ) $ becomes trivially
factorizable, since it is independent of $x$. A remarkable feature is
that the last two relations, Eq.~(\ref{eq:pd0}) and
Eq.~(\ref{eq:kp0}), are independent of the PV regulators. A similar
situation has also been encountered when computing PDF in the chiral
limit~\cite{DR95}; it was a constant equal to one, regardless on the
details of the PV regulator. We will show below that by putting together
Eq.~(\ref{eq:pd0}) and the results of Ref.~\cite{DR95} an interesting
relation follows.

Higher transverse moments diverge if one restricts the number of
Pauli-Villars subtractions to two, but Eq.~(\ref{eq:pd0}) and
Eq.~(\ref{eq:kp0}) remain still valid if more subtractions are
considered.

\begin{figure}[]
\includegraphics[width=8.5cm]{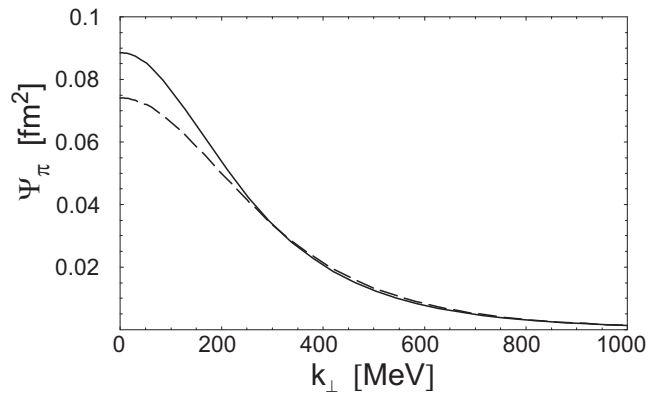} 
\caption{The pion light-cone wave function in the chiral limit,
evaluated in the Nambu--Jona-Lasinio with the Pauli-Villars
regularization with two subtractions and with the constituent quark
mass $M= 280 {\rm MeV}$ (solid line) and 350 MeV (dashed line), 
plotted as a function of the transverse
momentum $k_\perp$. The wave funcion does not depend on $x$.
The normalization is such that $ \int d^2 \vec k_\perp \Psi_\pi ( x ,
k_\perp )= \varphi_\pi (x) = 1 $. The second transverse moment is $
\langle \vec k_\perp^2 \rangle = -M \langle \bar u u \rangle
/f_\pi^2 = (625 {\rm MeV})^2$ for $M= 280 {\rm MeV}$ and $(634 {\rm MeV})^2$ 
for $M=350 {\rm MeV}$ . The scale relevant for the
calculation, as inferred from the QCD evolution \cite{DR95}, is $Q_0=313$~MeV. }
\label{fig:lcwf}
\end{figure}

In Fig.~(\ref{fig:lcwf}) we show the $k_\perp$-dependence of the
light-cone pion wave function in the chiral limit (finite pion mass
corrections turn out to be tiny, at the level of a few \%) for the PV regularization with two
subtractions, and with $M=380 {\rm MeV}$ and 350~MeV. For these values we get the
transverse moment $ \langle \vec k_\perp^2 \rangle = (625 {\rm
MeV})^2 $, and $(634 {\rm
MeV})^2 $, respectively. This value is about a factor of two  larger than the one
found in Ref.~\cite{PR01}, namely $(430{\rm MeV})^2$, and a factor of
four higher than the findings of Ref.~\cite{Zh94}, $(316 {\rm
MeV})^2$, at the scale at which $\alpha / \pi \sim 0.1 $, {\em i.e.}
$Q \sim 1-2 {\rm GeV}$. As we shall see below, a part of the
discrepancy can be attributed to the QCD radiative corrections.

In non-local versions of the chiral quark model, where a 
momentum-dependent mass function is introduced as a physically motivated
regulator, the trend to produce a constant PDA has also been observed
if the constant mass limit is considered~\cite{PP97,PP99,PR01}. In
those models such a limit effectively corresponds to removing the
regulator, against the original spirit of the model. Unfortunately,
for the genuine non-local case those calculations violate proper
normalization of PDA, because the employed
currents do not comply with the necessary Ward identities required by
chiral symmetry. The problem has been addressed in
Ref.~\cite{ADT01}, where it has been found that about a third of the
normalized PDA comes from the non-local currents. For a Gaussian mass
function there is a clear flattening of $\varphi_\pi (x)$ in the
central region of $0.2 \le x \le 0.8 $~\cite{Do02}.

We stress that our result, Eq.~(\ref{eq:pd0}), holds true without
removing the Pauli-Villars regulator and is in harmony with chiral
symmetry, since the starting point was the normal parity action, which
by construction preserves chiral symmetry. Obviously, the fact that
our final answer does not depend on the form of the PV regulators used makes
any subsequent manipulation with the regulators fully irrelevant.

Another point is that PDA from Eq. (\ref{eq:pd0}) and PDF from
Eq. (\ref{PDF1}) yield the relation $\varphi_\pi (x)= V_\pi (x)/2 $
valid at a low scale $Q_0$.  It is noteworthy that in the framework of
QCD sum rules the same identity between PDA and PDF has also been
obtained~\cite{BJ97} at some scale, although there the asymptotic form
for PDA was assumed without the QCD evolution, while PDF was obtained by
QCD evolution. We will show below that if evolution is undertaken for
both PDA and PDF at the same low energy scale, an overall consistent
picture arises.

\section{QCD evolution} 

The comparison of the leading-twist PDA to high-energy experimental
data requires, like for PDF, the inclusion of radiative logarithmic
corrections through the QCD evolution~\cite{BL79,Mu95}. For the pion
distribution amplitude this is done in terms of the Gegenbauer
polynomials, by interpreting our low-energy 
model result as the initial condition. For
clarity we work in the chiral limit, hence
\begin{eqnarray}
\varphi_{\pi} (x,Q_0) = 1.
\label{start}
\end{eqnarray}
Then, the LO-evolved distribution amplitude reads~\cite{BL79,Mu95}
\begin{eqnarray}
\varphi_{\pi} (x,Q) &=& 6x(1-x){\sum_{n=0}^\infty}'  C_n^{3/2} ( 2 x -1)
a_n (Q),
\label{eq:evolpda} 
\end{eqnarray}
where the prime indicates summation over even values of $n$ only. The matrix
elements, $a_n(Q)$, are the Gegenbauer moments given by
\begin{eqnarray}
a_n (Q)&=& \frac23 \frac{2n+3}{(n+1)(n+2)} \left(
 \frac{\alpha(Q_{})}{\alpha(Q_0) } \right)^{\gamma_n^{(0)} / (2
 \beta_0)} \times \nonumber \\ &&\int_0^1 dx C_n^{3/2} ( 2x -1)
 \varphi_{\pi} (x ,Q_0),
\label{Geg}
\end{eqnarray}
with $C_n^{3/2}$ denoting the Gegenbauer polynomials, and 
\begin{eqnarray}
\gamma_n^{(0)} &=& -2 C_F \left[ 3 + \frac{2}{(n+1)(n+2)}- 4
\sum_{k=1}^{n+1} \frac1k \right], \nonumber \\ 
\beta_0 &=& \frac{11}3 C_A -
\frac23 N_F,
\end{eqnarray}
with $C_A = 3$, $C_F = 4/3$, and $N_F$ being the number of active
flavors, which we take equal to three \footnote{The one-loop
anomalous dimension $\gamma_n^{(0)} > 0 $, and $\gamma_n^{(0)} \to 8
C_F \log n $ for large $n$, which coincides with the case of the
non-singlet parton distribution functions used in
Refs.~\cite{DR95,DR02}.}. With our constant amplitude (\ref{start}) we get immediately
\begin{eqnarray}
\int_0^1 dx C_n^{3/2} ( 2x -1) \varphi_{\pi} (x ,Q_0) =1.
\label{ourGeg}
\end{eqnarray} 
Thus, for a given value of $Q$ we may predict PDA.  
We need, however, to know what the initial scale $Q_0$ is,
or, equivalently, to know the evolution ratio 
$r=\alpha(Q) / \alpha(Q_0 )$. The fitting procedure of Ref.~\cite{SY00} yields 
$a_2 (2.4 {\rm GeV}) = 0.12 \pm 0.03$ (with the assumption $a_k=0$, $k > 2$). 
We treat this as experimental
input, and then with help of Eqs. (\ref{Geg},\ref{ourGeg})
we get for the evolution ratio
\begin{eqnarray}
\alpha(Q=2.4 {\rm GeV}) / \alpha(Q_0) = 0.15 \pm 0.06 .
\label{ourQ}
\end{eqnarray} 
which at LO implies $ Q_0 = 322 \pm 45 {\rm MeV}$, a value compatible within errors with 
(\ref{Q0}). 

The fit of Ref. \cite{SY00} with non-zero $a_4$ yields $a_2=0.19 \pm 0.04 \pm 0.09$ and
$a_4=-0.14 \pm 0.03 \mp 0.09$. The central value of $a_2$ would imply, according to our
prescription, the evolution ratio of $0.31$, and, correspondingly, 
$Q_0=0.47^{+0.51}_{-0.19}$~GeV, a much larger central value 
than (\ref{Q0}), but with very large errors. For that reason, in the numerical studies below we 
use the value (\ref{ourQ}) for the evolution ratio.

We can now predict the following lowest-order coefficients:
\begin{eqnarray}
&& a_4 (2.4 {\rm GeV}) = 0.044 \pm 0.016 \nonumber \\
% && \left( {\rm Ref. [5]: ~~~} 
% a_4=-0.14 \pm 0.03 \mp 0.09 \right) \nonumber \\ 
&& a_6 (2.4 {\rm GeV}) = 0.023 \pm 0.010 \nonumber \\ 
&& a_8 (2.4 {\rm GeV}) = 0.014 \pm 0.006 \\ 
&& a_{10} (2.4 {\rm GeV}) = 0.009 \pm 0.005 \nonumber 
%\\  \dots \nonumber 
\end{eqnarray} 
\begin{figure}[]
\includegraphics[width=8.5cm]{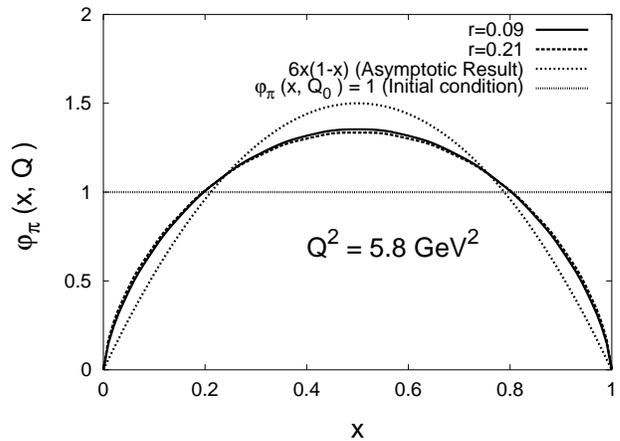}
\caption{The pion distribution amplitude in the chiral limit evolved to
the scale $Q^2 = (2.4 {\rm GeV})^2 $. 
The two values for the evolution ratio
$r=\alpha(Q) / \alpha(Q_0)$ reflect the uncertainties in the values of
Ref.~\cite{SY00} based on an analysis of the CLEO data.
We also show the unvolved PDA, $\varphi_\pi(x,Q_0)=1$, and the asymptotic PDA, 
$\varphi_\pi (x,\infty)=6x(1-x)$.}
\label{fig:pda}
\end{figure}
 
For the sum of the Gegenbauer
coefficients we get the estimate 
\begin{eqnarray}
{\sum_{n=2}^\infty}' a_n (Q=2.4{\rm GeV}) &=& \int_0^1 dx \frac{\varphi_\pi
(x,Q=2.4 {\rm GeV})}{6x(1-x)}-1  \nonumber \\ &=& 0.25 \pm 0.10
\label{eq:sumgeg} 
\end{eqnarray} 
where the uncertainties correspond to the uncertainties in
Eq.~(\ref{ourQ}). 

The leading-twist
contribution to the pion transition form factor is, at the LO in the QCD
evolution~\cite{BL80}, equal to   
\begin{eqnarray}
\frac{Q^2 F_{\gamma^* \to \pi \gamma} (Q) }{2 f_\pi} \Big|_{\rm twist-2} = 
\int_0^1 dx \frac{\varphi_\pi (x,Q )}{6x(1-x)}
\end{eqnarray}  
The experimental value obtained in CLEO~\cite{CLEO98} for the full
form factor is $ Q^2 F_{\gamma^* , \pi \gamma} (Q) / (2 f_\pi) = 0.83 \pm
0.12 $ at $Q^2 = {\rm (2.4 GeV)}^2 $. Our value for the integral, $1.25
\pm 0.10 $, overestimates the experimental result, although at
the $2 \sigma$-confidence level both numbers are compatible. Taking
into account the fact that we have not included neither NLO effects nor an
estimate of higher-twist contributions, the result is quite
encouraging.

In Fig.~\ref{fig:pda} we show our PDA evolved to $Q
= 2.4 {\rm GeV} $, for two values of the evolution ratio, which
reflect the uncertainties from Eq. (\ref{ourQ}). We also show the
initial and the asymptotic PDA's. It is interesting to note that after
evolution our results closely resemble those found in transverse
lattice approaches~\cite{Da01,BS01,BD02}. In particular, we get for
the second $\xi$-moment ($\xi = 2x-1$),
\begin{eqnarray}
\langle \xi^2 \rangle &=& \int_0^1 dx \, \varphi_\pi (x, Q=2.4 {\rm GeV} ) (2x-1)^2 
\nonumber \\ &=&
0.040 \pm 0.005,
\end{eqnarray} 
to be compared with $ \langle \xi^2 \rangle = 0.06 \pm 0.02 $ obtained
in the standard lattice QCD for $Q= 1/a = 2.6 \pm 0.1 {\rm
GeV}$~\cite{DP00}. From the PDF calculation at LO of Ref.~\cite{DR95}
we estimate that if the momentum fraction carried by the valence
quarks at $ Q = 2 {\rm GeV} $ is $0.47 \pm 0.02 \% $, then $Q_0$ is
such that $\alpha (Q_0)=2.14 $, and the evolution ratio at $Q=2 {\rm
GeV} $ is $r=0.15 $. Then, for $Q=2.4 {\rm GeV}$ we get $r=0.14$ from
the analysis of PDF, a value compatible, within uncertainties, with
the present calculation, Eq.~(\ref{ourQ}). This is a crucial finding,
showing the consistency of the results obtained in our approach. 

One might worry that the starting condition (\ref{start}) does not satisfy the
end-point vanishing behavior and therefore cannot be expanded in terms
of the Gegenbauer polynomials. This is true, provided one insists on
uniform pointwise convergence.  However, the Gegenbauer polynomials
form a complete set in the space of square-summable functions, hence
convergence may be understood in a weak sense \footnote{The condition
for $\varphi_\pi(x)$ to belong to such a space is $ \int_0^1 dx
\frac{\varphi_{\pi}(x)}{x(1-x)} < \infty$. The function
$\varphi_\pi(x) =1 $ does not belong to this space, but it belongs to
its closure. This resembles the well-known fact that plane waves do
not belong to the space of square-summable functions in the interval
$-\infty < x < \infty $, but nevertheless may be approximated by
square summable-functions.}.  The slow convergence is reflected by the
fact that in Fig.~\ref{fig:pda} at least 30-100 Gegenbauer polynomials
are needed for evolution ratios $r=0.9-0.21$ respectively. The convergence
at the mid-point, $x=1/2$, is improved, since the series for
$\varphi(x,Q)$ is sign-alternating. At the end-points, $ x= 0,1$, the
series diverges, since $ C_{2k}^{3/2}(\pm 1) =  \frac{1}{2} (2k+1)(2k+2)$, 
which means that the convergence in Eq. (\ref{eq:evolpda}) is not uniform.
In order to analyze the behavior close to the end-points in a greater detail 
we consider the large-$n$ contribution to Eq.~(\ref{eq:evolpda}). We have
\begin{eqnarray}
\left( \frac{\alpha(Q) }{\alpha(Q_0)} \right)^{\gamma_n^{(0)}/(2
\beta_0)} \to n^{- \frac{4 C_F}{ \beta_0} \ln
\frac{\alpha(Q_{})}{\alpha (Q_0)}},
\end{eqnarray}  
hence, for $Q \to Q_0 $,  $ Q > Q_0 $, and with $x \to 0$ (recall that the
function is symmetric under $x \to 1-x $), we obtain
\begin{eqnarray}
\varphi_{\pi} (x \to 0,Q ) & \to & 8 x \zeta \left( \frac{4 C_F}{2 \beta_0}
\ln \frac{\alpha(Q_{})}{\alpha (Q_0)} + 1 \right),
\end{eqnarray}  
where $\zeta(z) = \sum_{n=1}^\infty n^{-z}$ is the Riemann $\zeta$
function, and $\zeta(1) = \sum_{n=1}^\infty n^{-1} = \infty$. Thus the
slope of the evolved PDA at the end-points becomes steeper 
and steeper as $Q \to Q_0 $.

\begin{figure}[]
\includegraphics[width=8.5cm]{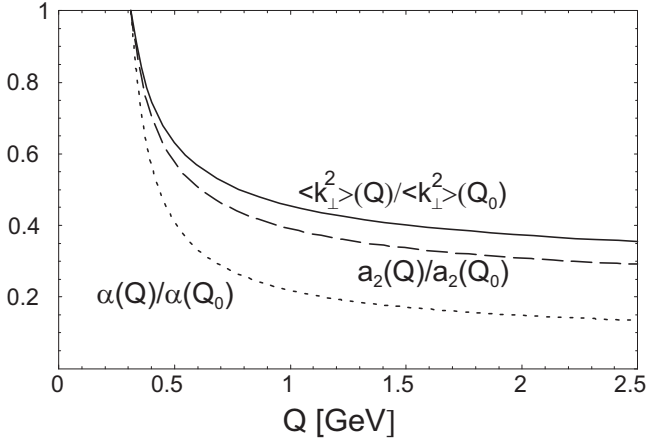}
\caption{Dependence of the
second transverse moment of the pion light-cone wave function, 
$ \langle k_\perp^2 \rangle_{Q_{}} / \langle k_\perp^2 \rangle_{Q_0} $ 
(solid line), the second Gegenbauer moment 
$ a_2 (Q)/a_2(Q_0) $ of the pion distribution amplitude (dashed line), and the evolution ratio 
$r=\alpha(Q)/\alpha(Q_0)$ (dotted line), plotted as functions of
the scale $Q$. The leading-order QCD evolution is applied. 
All quantities are relative to their values at the low energy scale,
$Q_0 = 313 \,{\rm MeV}$, at which the momentum fraction carried the quarks
equals unity~\cite{DR95}, according to the 
prescription that in a quark model $Q_0$ is defined by the
condition $ {\langle x V_\pi (x,Q_0) \rangle }= 1 $.  In our model $\alpha(Q_0)=2.14$,  
$a_2(Q_0) = 7/18$, and 
$ \langle k_\perp^2 \rangle_{Q_0} = (625 {\rm MeV})^2$ for $M= 280 {\rm MeV}$ and $(634 {\rm MeV})^2$ 
for $M=350 {\rm MeV}$ and in the chiral limit.}
\label{fig:evol}
\end{figure}

The QCD evolution also influences the value of the transverse moment.
According to the work of Ref.~\cite{Zh94}, $\langle \vec
k_\perp^2\rangle $ can be expressed as $\langle \vec
k_\perp^2\rangle = 5 m_0^2 / 36 $, where $m_0^2 = \langle \bar q
\sigma^{\mu \nu} F_{\mu \nu} q \rangle / \langle \bar q q \rangle $ is the
ratio between the quark-gluon and quark condensates. 
The quantity $m_0^2$ is scale dependent and
has been estimated to be $ m_0^2 (1 {\rm GeV}) = 0.8 \pm 0.2 {\rm
GeV}^2 $~\cite{BI82}.  Using the corresponding anomalous dimensions,
$4$ for $ \langle \bar q q \rangle$ and $-2/3$ for $\langle \bar q
\sigma^{\mu \nu} F_{\mu \nu} q \rangle $ \cite{VZS76}, yields
\begin{eqnarray}
\frac{\langle k_\perp^2 \rangle_{Q_{}} }{\langle k_\perp^2 \rangle_{Q_0} }
&=& \left( \frac{\alpha(Q_{})}{\alpha(Q_0)} \right)^{(4+2/3)/\beta_0}
\nonumber \\ &=& \left( \frac{\alpha(Q_{})}{\alpha(Q_0)} \right)^{14/(33-2N_f)}. 
\label{ratk}
\end{eqnarray} 
For $N_F=3$ this scale dependence can be seen in Fig.~(\ref{fig:evol}).
For the values $Q=1-2 \, {\rm GeV} $ one gets a reduction factor of $
0.37-0.45 $ for the ratio (\ref{ratk}), and $ \langle k_\perp^2 \rangle_{Q} = (430
{\rm MeV})^2 - (380 {\rm MeV})^2 $ for the second transverse moment, 
somewhat higher than the QCD sum rules estimate based on  
Ref.~\cite{Zh94},  $(316 {\rm MeV})^2$, or on Ref.~\cite{BI82},
$(333 \pm 40 {\rm MeV})^2$. 

\section{The relation to deep inelastic scattering} 

As we have already stated in Eq. (\ref{PDF1}), the valence PDF for the
pion in the chiral limit has also been found to be a constant equal to
one~\cite{DR95}.  At LO the non-singlet evolution of the PDF moments
is quite similar to that of the Gegenbauer moments of PDA,
Eq.~(\ref{eq:evolpda}), namely 
\begin{eqnarray} 
&& \int_0^1 dx \, x^n V_\pi (x,Q) =
\\ && \left( \frac{\alpha(Q_{})}{\alpha(Q_0)} \right)^{\gamma_n^{(0)}
/ (2 \beta_0)}  \int_0^1 dx \, x^n V_\pi (x,Q_0)= \nonumber \\ 
&& \frac2{n+1} \left( \frac{\alpha(Q_{})}{\alpha(Q_0)}
\right)^{\gamma_n^{(0)} / (2 \beta_0)}. \nonumber
\label{eq:evolpdf} 
\end{eqnarray} 
Thus, for $n=2$, one obtains 
\begin{eqnarray}
\frac{a_2(Q)}{a_2 (Q_0)} = \frac{\langle x^2 V_\pi (x,Q)
\rangle}{\langle x^2 V_\pi (x,Q_0)
\rangle} = \left( \frac{\alpha(Q_{})}{\alpha(Q_0)}
\right)^{\gamma_2^{(0)} / (2 \beta_0)}.
\end{eqnarray}
For $N_F=3$ this scale dependence for the ratios can be looked up in
Fig.~(\ref{fig:evol}). Using $ {\langle x^2 V_\pi (x,Q_0) \rangle }= 2/3 $
and $a_2(Q_0)=7/18 $ yields
\begin{eqnarray}
\frac{ a_2 (Q) }{\langle x^2 V_\pi (x,Q) \rangle} &=& \frac{7}{12},
\end{eqnarray}
hence $a_2 ( 2 {\rm GeV}) = 0.12 \pm 0.01 $ for $\langle x^2 V_\pi
\rangle = 0.20 \pm 0.01 $~\cite{SM92} and $a_2 ( 2 {\rm GeV} ) = 0.10
\pm 0.01 $ for $\langle x^2 V_\pi \rangle = 0.17 \pm 0.01
$~\cite{GRV99}. 

One can combine
Eqs.~(\ref{PDF1},\ref{eq:pd0},\ref{eq:evolpda},\ref{eq:evolpdf}) to
obtain the following very interesting LO relation that holds in the
considered model:
\begin{eqnarray}
\frac{\varphi_\pi (x,Q)}{6x(1-x)}-1 = \int_0^1 dy K(x,y) V_\pi ( y, Q )  ,
\label{eq:evol}
\end{eqnarray} 
where the kernel $K$ is independent of $Q^2$, and is given by
\begin{eqnarray}
K(x,y) = {\sum_{n=2}^\infty}' \frac{(2n+3)}{3(n+2)} C_n^{3/2}
(2x-1) y^{n}.
\end{eqnarray}  
In general, the relation (\ref{eq:evol}) holds in any model where PDA
and PDF are simultaneously equal to unity at some scale $Q_0$, and are
subsequently evolved at LO. Physically, Eq.~(\ref{eq:evol}) simply
tells us that the departure of PDA at a given $Q^2$ from the
asymptotic form is proportional to a weighted integral of PDF at the
same $Q$.  Clearly, $\varphi_\pi(x,Q) \to 6 x (1-x) $ if $ V_\pi (x,Q
) \to 2 \delta (x) $ or, equivalently, $ x V_\pi (x,Q) \to 0$, since
$K(x,0)=0$. Roughly speaking, in the present model the pion
distribution function is as close to the asymptotic value as the
non-singlet parton distribution. A remarkable feature of relation
(\ref{eq:evol}) is that it binds matrix elements related to exclusive
(PDA) and to inclusive (PDF) processes.

In order to evaluate the kernel we use the
symmetrized generating function of the Gegenbauer polynomials,
\begin{eqnarray}
G(x,y) &=& {\sum_{n=2}^\infty}' C_n^{3/2} (2x-1) y^n = \frac12
\left\{ R_+^{-3/2} + R_-^{-3/2} \right\} -1 \nonumber \\ R_\pm &=& 1 \mp
2(2x-1)y + y^2, 
\end{eqnarray} 
whence one can obtain 
\begin{eqnarray}
K(x,y) = \frac2{3} G(x,y) - \frac1{3y^2} \int_0^y d y' y' G(x,y').
\end{eqnarray} 
The integrals can be worked out to yield the final result 
\begin{eqnarray}
&& K(x,y) = \frac{1}{24 R_+^{3/2} y^2 (x-1)x}\times \nonumber \\&&
[8\left( x-1 \right) x y^2 + 
  R_+\left( \left(  2x-1 \right) y-1
     \right) \nonumber \\ && + 2{\sqrt{R_+}}
   \left( x-1 \right) x y^2
   \left( 1 + \left( 2 - 4x \right) y + 
     y^2 \right) \nonumber \\ &&  + R_+^{\frac{3}{2}}
   \left( 1 - 8 \left( x-1 \right) x
      y^2 \right)] - \quad \left( y \leftrightarrow - y \right)
\end{eqnarray} 
To test the success of Eq.~(\ref{eq:evol}) we need some input for
$V_\pi(x,Q ) $. However, taking into account the fact that the
agreement of the evolved valence PDF, $V_\pi(x,Q ) $ with the
parameterization of Ref.~\cite{SM92} at $Q^2 = 4 {\rm GeV}^2 $ is
almost perfect~\cite{DR95,DR02}, and that the results are almost
insensitive to the evolution ratio, $ \alpha(Q )/ \alpha(Q_0 ) $,
Fig.~\ref{fig:pda} can be regarded as a direct prediction of
Eq.~(\ref{eq:evol}) taking Ref.~\cite{SM92} as input for $ V_\pi
(x,Q)$. A further consequence of Eq.~(\ref{eq:evol}) may be obtained
by integrating with respect to $x$ and performing the sum over
$n$. Through the use of Eq.~(\ref{eq:sumgeg}) we  get 
\begin{eqnarray}
{\sum_{n=2}^\infty}' a_n (Q) = \int_0^1 dy \kappa(y) V_\pi (y,Q) \, 
\end{eqnarray} 
where 
\begin{eqnarray}
\kappa(y)&=& \int_0^1  dx K(x,y) = {\sum_{n=2}^\infty}' \frac{(2n+3)}{3(n+2)} y^{n} \\
&=&  \frac{3y^2+1}{6(1-y^2)} +\frac{\log (1 - y)  + \log (1 + y) }{6y^2}  \nonumber .  
\end{eqnarray} 
Notice that, for $Q \to \infty $ we get $ V_\pi (x,Q) \to 2 \delta (x)
$ and since $\kappa (y) = 7y^2 /12 + {\cal O} (y^4) $ one gets
${\sum_{n=2}^\infty }' a_n (Q) \to 0$, as expected.  Finally, using
the parameterization of Ref.~\cite{SM92} we get~\footnote{This is $ x
V_\pi (x,Q) = A_V x^\alpha (1-x)^\beta $ with $A_V$ such that $
\langle V_\pi \rangle = 2 $ and $\alpha = 0.64 \pm 0.03 $ and $
\beta=1.08 \pm 0.02 $ (the NA10 set) and $ \beta = 1.15 \pm 0.02 $
(the E615 set). Our estimate of error includes both sets.}
\begin{equation} 
{\sum_{n=2}^{\infty}}' a_n (2 {\rm GeV}) = 0.25 \pm 0.03 ,
\end{equation} 
a value perfectly compatible with Eq.~(\ref{eq:sumgeg}) although with smaller
uncertainties~\footnote{Of course, this estimate does not include
systematic uncertainties in NLO both for PDA and PDF.}. Again, this
verifies the consistency of our approach.

\section{Conclusions} 
We summarize our points. We have computed the light-cone pion wave
function and the pion distribution amplitude in the
Nambu--Jona-Lasinio model.  To this end, and to comply with previous
results regarding the parton distribution functions, we have used the
Pauli-Villars regularization method in such a way that chiral
symmetry, gauge invariance, and relativistic invariance are
preserved. As a result, we find that in the chiral limit the pion
distribution amplitude, computed as a low energy matrix element of an
appropriate operator, is a constant equal to one, $ \varphi_\pi(x)=1$,
and the second transverse moment of the pion light-cone wave function
is $\langle \vec k_\perp^2 \rangle = -M \langle \bar u u \rangle
/f_\pi^2 $, with $M$ denoting the constituent quark mass.  Both
results are independent of the particular form of the Pauli-Villars
regulators used. After the QCD evolution of the pion distribution
amplitude to the experimentally accessible region we find a result
still rather far away from the asymptotic form, $\varphi_\pi (x) = 6 x
(1-x) $, but in a good agreement with the analysis of the experimental
data from the CLEO collaboration. We can determine the working
momentum scale for the model to be $Q_0=313$~MeV, a rather low
value. Moreover, the scale $Q_0$ obtained in this work is compatible,
within experimental uncertainties, to the value obtained from the
previous analysis of the parton distribution functions, carried out
within exactly the same model. At the scale $Q_0$ the quarks carry all
the momentum of the pion. Our value obtained for the second transverse
moment of the pion light-cone wave function, $\langle \vec k_\perp^2
\rangle $, becomes, after the QCD evolution, not far from the
estimates based on the QCD sum rules. Finally, we have also derived a
model relation which binds the departure of the pion distribution
amplitude from its asymptotic value to an integral involving the pion
quark distribution function.  The relation, specific to the feature of
our model that at the scale $Q_0$ both the PDA and PDF are  constant
and equal to unity, has been successfully checked against the
available data.

\vskip0.5cm

We thank Micha\l{} Prasza\l{}owicz for several stimulating
discussions. This work has been partially supported by the DGES under
contract PB98-1367 and by the Junta de Andaluc{\'{\i}}a (Spain). Partial 
support from the Spanish Ministerio de Asuntos Exteriores and the Polish
State Committee for Scientific Research, grant number 07/2001-2002, is also
gratefully acknowledged.

\end{document}